\documentclass[letter]{aa} % for the letters 

\bibpunct{(}{)}{;}{a}{}{,}

\usepackage{graphicx}
\usepackage{txfonts}
\begin{document}

   \title{Excellent daytime seeing at Dome Fuji on the Antarctic plateau}

   \author{H. Okita
          \inst{1}
          \and
          T. Ichikawa
          \inst{1}
          \and
          M.C.B. Ashley
          \inst{2}
          \and
          N. Takato
          \inst{3}
          }
             
   \institute{Astronomical Institute, Tohoku University,
              6-3 Aramaki, Aoba-ku Sendai 980-8578, Japan\\
              \email{h-okita@astr.tohoku.ac.jp}
              \and
              School of Physics, University of New South Wales, Sydney, NSW 2052, Australia
              \and
              Subaru Telescope, 650 North A`ohoku Place, Hilo, HI 96720, USA
             }

   \date{Received May 22, 2013; accepted MM DD, YYYY}

% \abstract{}{}{}{}{} 
% 5 {} token are mandatory
 
  \abstract
  % context heading (optional)
  % {} leave it empty if necessary  
   {Dome Fuji, the second highest region on the Antarctic plateau, is expected to have some of the best astronomical seeing on Earth. However, site testing at Dome Fuji is still in its very early stages.}
  % aims heading (mandatory)
   {To investigate the astronomical seeing in the free atmosphere above Dome Fuji, and to determine the height of the surface boundary layer.}
  % methods heading (mandatory)
   {A Differential Image Motion Monitor was used to measure the seeing in the visible (472 nm) at a height of 11 m above the snow surface at Dome Fuji during the austral summer of 2012/2013.}
  % results heading (mandatory)
   {Seeing below 0.2$^{\prime\prime}$ has been observed.
The seeing often has a local minimum of $\sim$ 0.3$^{\prime\prime}$ near 18 h local time.
Some periods of excellent seeing, 0.3$^{\prime\prime}$ or smaller, were also observed, sometimes extending for several hours at local midnight.
The median seeing is higher, at 0.52$^{\prime\prime}$---this large value is believed to be caused by periods when the telescope was within the turbulent boundary layer.}
  % conclusions heading (optional), leave it empty if necessary 
   {The diurnal variation of the daytime seeing at Dome Fuji is similar to that reported for Dome C, and the height of the surface boundary layer is consistent with previous simulations for Dome Fuji.
The free atmosphere seeing is $\sim$ 0.2$^{\prime\prime}$, and the height of the surface boundary layer can be as low as
 $\sim$ 11 m.}


   \maketitle
%
%________________________________________________________________

%####################################################################
% Introduction
%####################################################################
\section{Introduction}
Dome Fuji is located at 77$^\circ$19$^\prime$S\ \ 39$^\circ$42$^\prime$E and, with a height of 3\,810 m, is the second highest region on the Antarctic plateau.
Astronomical seeing is generally considered as the superposition of the contributions from two layers; 
the surface boundary layer and the free atmosphere above.
Measurements at Dome C, which is another high region on the Antarctic plateau, have shown the best seeing so far observed from the Earth with a free atmosphere seeing of $\sim$ 0.3$^{\prime\prime}$ and a surface boundary layer thickness of $\sim$ 30 m \citep{lawrence2004,aristidi2009}.
%%For Dome A, which is the highest region of the Antarctic plateau, the values are 0.22$^{\prime\prime}$ and 14 m \citep{saunders2009,bonner2010}. 
At Dome A, the highest region on the Antarctic plateau, the median height of the surface boundary layer has been  measured as $\sim$14 m \citep{bonner2010}.
Simulations suggest that the free atmosphere seeing above Dome Fuji and the height of the surface boundary layer could be 0.21$^{\prime\prime}$ and 18 m \citep{saunders2009,sg2006}.
We conducted a site-testing campaign over the 2012/2013 summer in an attempt to test these expectations.

%####################################################################
% Instrumentation
%####################################################################
\section{Instrumentation}
The Differential Image Motion Monitor (DIMM) is a commonly-used instrument to measure the seeing.
The DIMM works by using two sub-apertures on a small telescope, with a wedge prism attached, to make two images of the same star on a CCD detector. 
By measuring the differential motion between these two images and assuming Kolmogorov turbulence, the seeing can be calculated \citep{sr1990}.

\begin{figure}
	\resizebox{\hsize}{!}{\includegraphics[angle=-90]{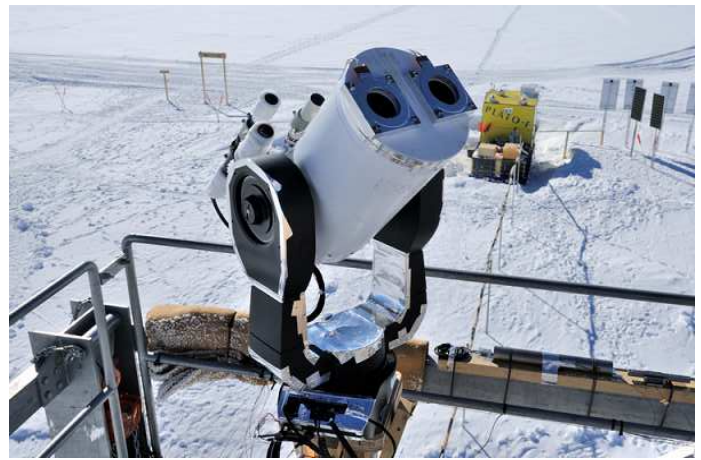}}
	\caption{DF-DIMM on the 9 m tower at Dome Fuji, 2013 January. The telescope aperture is $\sim$11 m above the snow surface.}
	\label{fig00}
\end{figure}

Our instrument, the ``Dome Fuji Differential Image Motion Monitor'' (DF-DIMM), is based on a Meade LX200-8'' cassegrain telescope, with an SBIG ST-i monochrome CCD camera equipped with an Edmund narrow-band filter at 472 nm for suppressing auroral emissions \citep{okita2013}.
DF-DIMM (see Figure \ref{fig00}) was operated fully automatically to allow the efficient accumulation of many seeing estimates.
The optical tube of the telescope was painted white to minimize the local turbulence inside and around the tube generated by the solar radiation.
Many modifications were made to allow operation in the Antarctic environment.
Table \ref{dfdimm.table} summarizes the parameters and the technical specifications of DF-DIMM.
\begin{table}
	\caption{Parameters and technical specifications of DF-DIMM.}
	\label{dfdimm.table} 
	\centering
	\begin{tabular}{l l}
	\hline \hline
	Sub-aperture diameter            & $\phi$ 60 mm                   \\
	Sub-aperture separation          & 140 mm                         \\
    Observed wavelength              & 472 nm                         \\
    FWHM of the filter               & 35 nm                          \\
    Pixel size                       & 7.4 $\mu$m $\times$ 7.4 $\mu$m \\
    Pixel scale                      & 0.775$^{\prime\prime}$/pix $\pm$0.005$^{\prime\prime}$/pix \\
    Exposure time                    & 0.001 sec                      \\
    Number of frames used            &                                \\
    \qquad for each seeing estimate  & 450 over $\sim$5 minutes       \\
	Height of the entrance pupils    & $\sim$11 m                     \\
	\hline
	\end{tabular}
\end{table}

DF-DIMM observed Canopus ($\alpha$ Car, V = $-$0.7 mag, the second brightest star in the sky) to measure the seeing.
Canopus is circumpolar at Dome Fuji, with a zenith angle varying from 25$^\circ$ to 50$^\circ$.
DF-DIMM could observe Canopus continuously for days at a time with a reasonable contrast against the daytime sky background.

DF-DIMM was placed on the top of a 9 m tower in order to be as high as possible within, and sometimes above, the surface boundary layer.
The height of the entrance pupils of DF-DIMM was $\sim$ 11 m. 

\begin{figure}
	\resizebox{\hsize}{!}{\includegraphics[angle=-90]{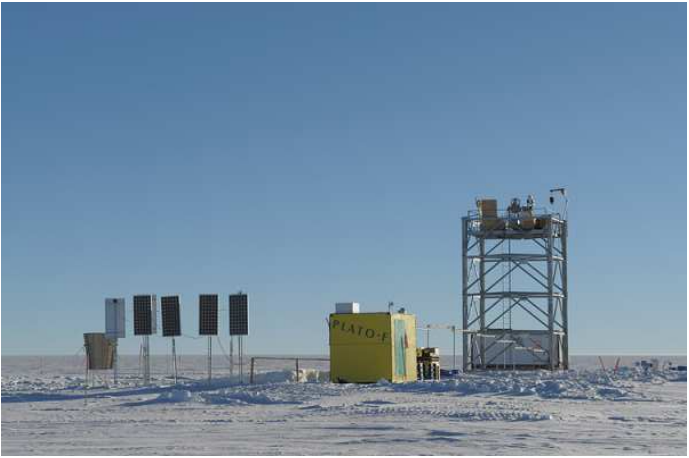}}
	\caption{PLATO-F (center, yellow container) and the 9 m tower (right) with DF-DIMM on top. The rectangular objects to the left are solar panels.}
	\label{fig01}
\end{figure}

DF-DIMM was supported by PLATO-F (PLATeau Observatory for Dome Fuji), a fully automated observing platform for the Antarctic plateau deployed at Dome Fuji in January 2011.
Figure \ref{fig01} shows PLATO-F and the 9 m tower.
PLATO-F provides electrical power of up to 1 kW and Iridium communications all year for site testings and astronomical observations from Dome Fuji \citep{ashley2010}.

%####################################################################
% Data processing and error analysis
%####################################################################
\section{Data processing and error analysis}
On-site data processing software was developed for DF-DIMM, based on Nightview\footnote{http://www.physics.muni.cz/mb/nightview/nightview.html} (Hroch), Sextractor \citep{sextractor}, and CFITSIO \citep{cfitsio}.
The longitudinal seeing $\epsilon_l$ and the transverse seeing $\epsilon_t$ were calculated using the equations (13), (14), and (23) of \citet{sr1990} from 450 images taken at about five-minute intervals.
The seeing estimates were then corrected for zenith angle using equation (24) of the paper.

The seeing values measured by a DIMM, $\epsilon_l$ and $\epsilon_t$, have statistical error and pixel scale uncertainty.
The statistical error of the variance of a star position $\sigma_\star^2$ is   
$d\sigma_\star^2/\sigma_\star^2$=$\sqrt{\ 2 /(N-1)}$, where $N$ is the number of frames used for the variance calculation and the subscript $\star$ represents either longitudinal or transverse \citep{sr1990}.
In our case, 450 frames were used in calculating each seeing estimate.
Hence, the statistical error on the variance is $\sim$ 6.7\%, which corresponds to a seeing error of 
$d\epsilon_\star/\epsilon_\star\propto(d\sigma_\star^2/\sigma_\star^2)^{3/5}\sim$ 4\%.
The pixel scale of DF-DIMM was measured using the diurnal motion of Canopus on January 1.
This gives a scale of 0.775$^{\prime\prime}$ $\pm$ 0.005$^{\prime\prime}$ per pixel.
The uncertainty contributes $\sim$ 0.8\% error in the seeing.
The temperature dependence of the focal length also affects the pixel scale.
However this effect is negligibly small for our Cassegrain telescope.
In fact, optical simulations demonstrate that the focal length of DF-DIMM changes less than 0.2\% between 
$20^\circ$C and $-80^\circ$C.

%%Seeing should be the same in the longitudinal and transverse directions because Kolmogorov turbulence with infinite outer scale is assumed. 
Seeing should be the same in the longitudinal and transverse directions since seeing is a scalar quantity. 
Considering the statistical error and pixel scale uncertainty, we discarded $\sim$1 \% of observations that fell outside the range $0.50<\epsilon_l/\epsilon_t<2.0$.
We then averaged $\epsilon_l$ and $\epsilon_t$ to obtain the seeing value.

Other sources of uncertainty in the DIMM measurements are considered below for better estimation of seeing.

\subsection{Instrument rotation effect}
\begin{figure}
	\resizebox{\hsize}{!}{\includegraphics{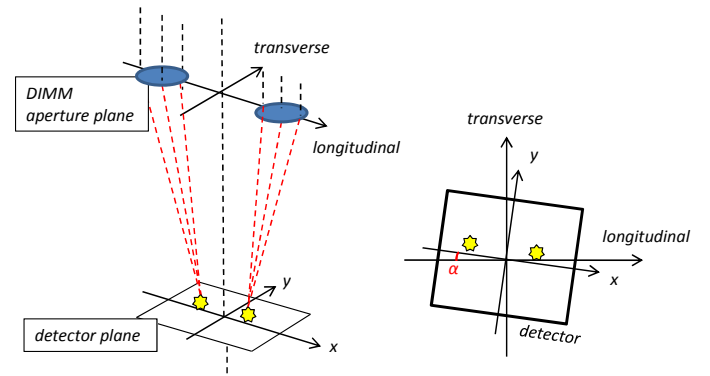}}
	\caption{Schematic showing the ray trace for a generic DIMM. Two sub-apertures with wedge prisms make two images of the same star on a CCD detector. 
%%Here the detector plane of ($x$, $y$) is rotated an angle of $\alpha$ with respect to the DIMM aperture plane 
%%of ($longitudinal$, $transverse$), the IRE occurs.
    }
	\label{ire}
\end{figure}

To simplify the analysis of DIMM data, it is usual to align the ($x$, $y$) coordinates of the CCD detector with the longitudinal and transverse DIMM coordinates ($l$, $t$) defined by \citet{sr1990}, see Fig.\ref{ire}. If, however, these coordinate frames are misaligned by some angle $\alpha$, and this is not corrected for in the analysis, an error will result.

Here we write $\sigma_x^2$ and $\sigma_y^2$ as the variance of the differential motion along the $x$ and $y$ axes, 
and $\sigma_{xy}$ as the covariance of $x$ and $y$.
The longitudinal variances of $\sigma_l^2$ and transverse variance of $\sigma_t^2$ are then
\begin{equation}
\sigma_l^2
=\cos^2(\alpha)\sigma_x^2
+\sin^2(\alpha)\sigma_y^2
-\sin(2\alpha)\sigma_{xy}
\end{equation}
\begin{equation}
\sigma_t^2
=\sin^2(\alpha)\sigma_x^2
+\cos^2(\alpha)\sigma_y^2
+\sin(2\alpha)\sigma_{xy}\ .
\end{equation}

For precise measurements of the seeing from DIMM observations we need to transform the ($x$, $y$) coordinates to ($l$, $t$) before using the normal DIMM equations.

\subsection{Finite exposure effect}
Theoretically, DIMM seeing is defined in an infinitely short exposure.
\citet{martin1987} and \citet{soules1996} discussed the effect of using a finite exposure time.
From equation (18) of \citet{soules1996} with $\tau$=0.001 s and $w\le$30 m$/$s, our seeing values are underestimated by less than 3 \%.

\subsection{Readout, background, and local turbulence effects}
Readout noise and background noise of the detector also add small biases to the seeing value \citep{tokovinin2002}.
The local turbulence inside the telescope worsens the observed seeing.
These effects all cause our results to be an upper limit on the actual seeing.
%%\begin{eqnarray}
%%\sigma_{apparent\ \ast}^2
%%=\sigma_{true\ \ast}^2
%%+\sigma_{background\ \ast}^2
%%+\sigma_{readout\ \ast}^2
%%+\sigma_{telescope\ \ast}^2
%%\end{eqnarray}
%%The subscript $\ast$ represent either $x$ or $y$.

%####################################################################
% Results
%####################################################################
\section{Results}
\begin{figure*}
	\centering
	\includegraphics[width=17cm, angle=0]{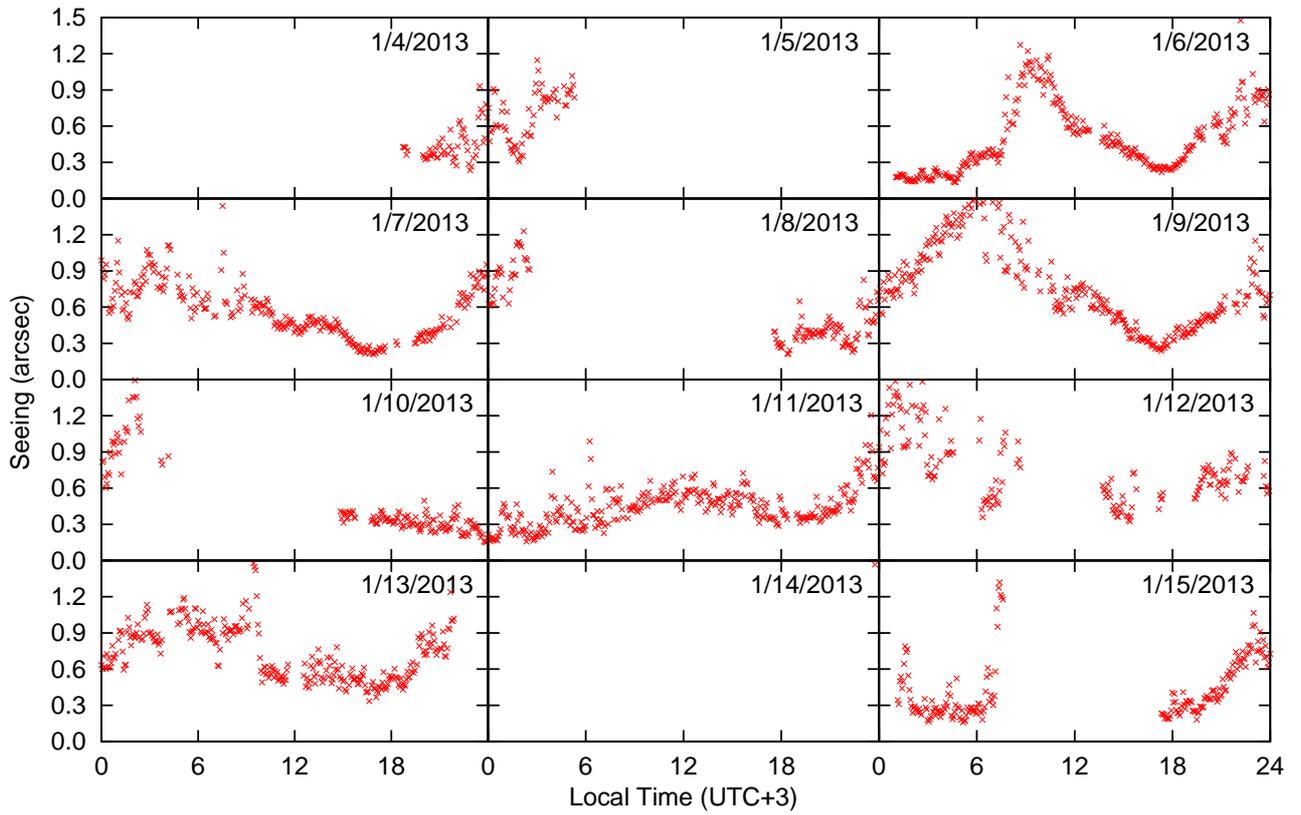}
	\caption{Time series of Dome Fuji seeing from 2013 January 4 to 15. The seeing was measured at wavelength of 472 nm at a height of 11 m above the snow surface. We plot the average of the longitudinal and transverse seeings.}
	\label{fig1}
\end{figure*}

\begin{figure*}	
	\centering
	\includegraphics[width=17cm, angle=0]{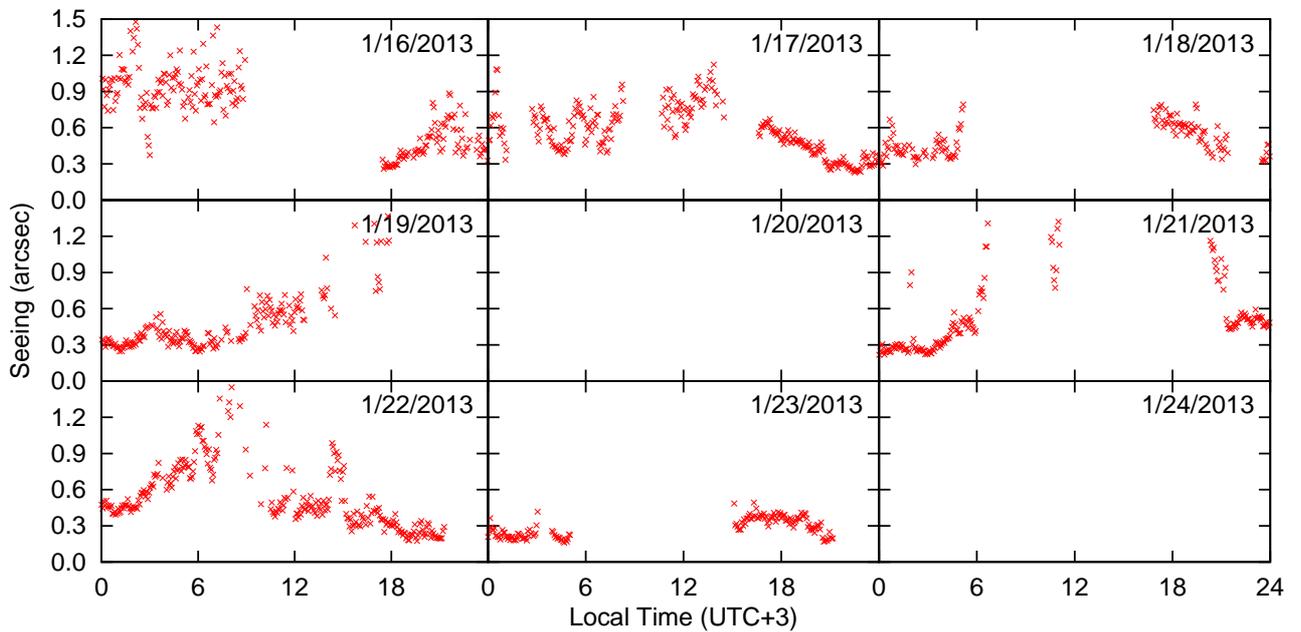}
	\caption{Same as Fig. \ref{fig1}, but for the period 2013 January 16 to 24.}
	\label{fig2}
\end{figure*}

We carried out DIMM observations 11 m above the snow surface at a wavelength of 472 nm from 2013 January 4 to January 23.
In all, we obtained 3\,814 seeing estimates, each one calculated from 450 images over a period of about five-minutes.
Figure \ref{fig1} and \ref{fig2} show the time series of the seeing, day by day.

A period of excellent seeing, below 0.2$^{\prime\prime}$ and continuing for about 4 hours, was observed near local midnight on 2013 January 6.
Other periods of excellent seeing, less than 0.3$^{\prime\prime}$, were observed close to local midnight on a total of six occasions (January 6, 11, 15, 19, 21 and 23).

The seeing has a tendency to have a local minimum of $\sim$ 0.3$^{\prime\prime}$ near 18 h local time.
This is clear in the data for January 6, 7, 9, and 16.

The histogram of the seeing measurements is plotted in Figure \ref{fig3}.
The mean, median, and mode of the seeing values were 0.68$^{\prime\prime}$, 0.52$^{\prime\prime}$, and 0.36$^{\prime\prime}$, respectively.
The 25th and the 75th percentile of seeing were 0.36$^{\prime\prime}$ and 0.78$^{\prime\prime}$.
As discussed below, we expect the higher seeing measurements to be due to periods when the surface boundary layer was above the level of the top of the telescope.

%####################################################################
% Discussion and conclusion
%####################################################################
\section{Discussion and conclusion}
We note that the Dome Fuji seeing tends to have its smallest values a few hours around local dusk and midnight.
It is remarkable that seeing in the range 0.2$^{\prime\prime}$ to 0.3$^{\prime\prime}$ was observed for continuous periods of hours at a height of only 11 m above the snow surface, presumably due to periods where the surface boundary layer is either below the height of the DIMM aperture, or has disappeared altogether.

A similar local minimum at local dusk has also been seen at Dome C, and has been interpreted by 
\citet{aristidi2005} as due to the disappearance of the surface boundary layer.
Our results are consistent with this.

However, it is interesting to note that the excellent seeing we have observed at local midnight has not been reported from site testing of 8 m above the snow surface at Dome C. 
The weak insolation at midnight is expected to result in an intense temperature gradient near the snow surface at this time.
This strong temperature gradient should produce a strong surface boundary layer, and hence poor seeing from the surface. 
This is only consistent with our observations if the surface boundary layer is below the level of our telescope. 
We therefore conclude that our DIMM was above the surface boundary layer during these periods, and was sampling the free atmosphere seeing.
A low surface boundary layer has been predicted from simulations by \citet{sg2006}, and is consistent with our observations.
Observations with a sonic radar at Dome A have shown that the surface boundary layer is often highly turbulent, but confined to a very thin (14 m median) layer near the snow \citep{bonner2010}.

\begin{figure}
	\resizebox{\hsize}{!}{\includegraphics[angle=0]{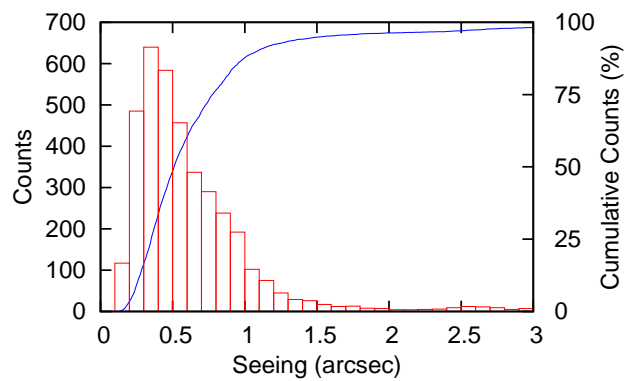}}
	\caption{Histogram (bars) and cumulative histogram (solid line) of Dome Fuji seeing measured from 2013 January 4 to 23. Note that measurements above $\sim$ 0.4$^{\prime\prime}$ are likely the result of the DIMM being within the surface boundary layer.}
	\label{fig3}
\end{figure}

The histogram of seeing measurements in Figure \ref{fig3} is expected to consist of two sets of data: those when the telescope is outside the surface boundary layer, and those when the telescope is inside. The latter measurements will produce the long tail of seeing measurements above $\sim$ 0.4$^{\prime\prime}$. We expect that if the DIMM was mounted on a higher tower, the fraction of measurements in this tail would drop significantly. A quantitative estimate of this effect could be made from sonic radar measurements of the boundary layer height.

In summary, at a given height above the snow, excellent seeing at Dome Fuji occurs when there is either a low or non-existent surface boundary layer.
The free atmosphere seeing is $\sim$ 0.2$^{\prime\prime}$.
The height of the surface boundary layer has been observed to be as low as $\sim$ 11 m.

Our findings give strong encouragement to constructing future large-aperture telescopes on the Antarctic plateau to take advantage of the excellent natural seeing and the low surface boundary layer.
We are now preparing to make wintertime seeing measurements with DF-DIMM.

%####################################################################
% aknowledgements
%####################################################################

\begin{acknowledgements}
We acknowledge the National Institute of Polar Research and the 51st - 54th Japanese Antarctic Research Expeditions.
This research is supported by the National Institute of Polar Research through Project Research no.KP-12, the Grants-in-Aid for Scientific Research 18340050 and 23103002, the Australian Research Council and Australian government infrastructure funding managed by Astronomy Australia Limited. 
Hirofumi Okita thanks the Sasakawa Scientific Research Grant from The Japan Science Society, and Tohoku University International Advanced Research and Education Organization for scholarships and research expenses.
\end{acknowledgements}

%####################################################################
% bibliography
%####################################################################

\bibliographystyle{aa.bst}
\bibliography{okita.bib}

\end{document}